\documentclass[12pt]{article}
\pdfoutput=1
\usepackage[utf8]{inputenc}
\usepackage{lutypaper,graphicx,xcolor,appendix}
\usepackage{epsfig}
\usepackage{graphicx}
\usepackage[export]{adjustbox}
\usepackage{multirow}
\usepackage{slashed}
\numberwithin{equation}{section} % Equation numbering within sections

\renewcommand{\tilde}{\widetilde}
\renewcommand{\bar}{\widebar}
\renewcommand{\O}{\scr{O}}

\begin{document}

\newcommand{\1}{\mathds{1}}
\newcommand{\lla}{\langle\!\!\ggap\langle}
\newcommand{\rra}{\rangle\!\!\ggap\rangle}

% =============================================================================
% Title page
% =============================================================================
\begin{titlepage}

\title{Convergent Momentum-Space OPE\\
and Bootstrap Equations in\\
Conformal Field Theory}

\author{Marc Gillioz$^{a,b}$, Xiaochuan Lu$^c$, Markus A.~Luty$^d$, and Guram Mikaberidze$^d$}
\address{$^a$SISSA, via Bonomea 265, 34136 Trieste, Italy}
\address{$^b$Theoretical Particle Physics Laboratory,
Institute of Physics \\
EPFL, Lausanne, Switzerland}
\address{$^c$Institute of Theoretical Science \\
Department of Physics \\
University of Oregon, Eugene, OR 97403, USA}
\address{$^d$Center for Quantum Mathematics and Physics (QMAP)\\
University of California, Davis, California 95616}

\vspace{10mm}

\begin{abstract}
General principles of quantum field theory imply that there exists an operator
product expansion (OPE) for Wightman functions in Minkowski momentum space that converges
for arbitrary kinematics.
This convergence is guaranteed
to hold in the sense of a distribution,
meaning that it holds for correlation functions smeared by smooth test functions.
The conformal blocks for this OPE are conceptually extremely simple: they
are products of 3-point functions.
We construct the conformal blocks in 2-dimensional conformal field theory and show
that the OPE in fact converges pointwise to an ordinary function
in a specific kinematic region.
Using microcausality, we also formulate a bootstrap equation directly
in terms of momentum space Wightman functions.

\end{abstract}

\end{titlepage}

\tableofcontents

\noindent

% =============================================================================
\section{Introduction}
\label{sec:Intro}
% =============================================================================

In this paper we study
the operator product expansion (OPE)
in general conformal field theories (CFTs) in momentum space, and
formulate a bootstrap equation for the CFT data.
There have been a number of studies of aspects of CFT in momentum space,
including formal developments
\cite{Coriano:2013jba, Bzowski:2013sza, Isono:2019ihz, Bautista:2019qxj, Gillioz:2019lgs},
but also following physical motivations, such as the study of anomalies~\cite{Bzowski:2015pba, Coriano:2017mux, Gillioz:2016jnn, Gillioz:2018kwh},
the formulation of Hamiltonian truncation~\cite{Katz:2016hxp, Fitzpatrick:2018ttk, Anand:2019lkt},
and early-universe cosmology~\cite{Arkani-Hamed:2018kmz, Sleight:2019hfp, Baumann:2019oyu}.
Correlation functions in momentum space exist as the Fourier transform of correlation functions in position space.
However, time-ordered correlation functions in momentum space are not expected to have a convergent OPE.
(This includes Euclidean correlation functions, which are radially
ordered in the operator formalism.)
The reason is that when the OPE converges,
it corresponds to an insertion of a complete set of states in some quantization.
The Fourier transform involves an integral over all possible positions of the operators,
and hence all possible time orderings.
Therefore the Fourier transform of time-ordered correlators
cannot be written as a vacuum expectation value of a product of operators in
any simple sense.
For this reason, we focus on Wightman functions, products of operators with
fixed ordering:
\[
\eql{Wightmandefn}
\bra{0} \tilde{\scr{O}}_n(p_n) \cdots \tilde{\scr{O}}_1(p_1) \ket{0}
&\equiv (2\pi)^d \de^d(p_n + \cdots + p_1) \lla
\tilde{\scr{O}}_n(p_n) \cdots \tilde{\scr{O}}_1(p_1) \rra.
\]
These correlation functions are well-defined in Minkowski space, but not in
Euclidean space.
The reason is that in Euclidean space, correlation functions of operators
do not make sense if the operators are out of time order.
For example, for the conventional quantization where $x^0$ is the time variable,
we have
\[
\bra{0} &\scr{O}_n(x_n) \ggap \cdots  \scr{O}_1(x_1)\ket{0}_\text{E}
\nn
&\ \ {}= \bra{0} \scr{O}_n(0, \vec{x}_n) e^{-H (x_n^0 - x_{n-1}^0)}
\scr{O}_{n-1}(0, \vec{x}_{n-1}) \cdots \scr{O}_2(0, \vec{x}_2)
e^{-H (x_2^0 - x_1^0)} \scr{O}_1(0, \vec{x}_1) \ket{0}_\text{E}.
\]
Unless the time differences in the exponents are all positive, the time evolution
operators are
ill-defined because the Hamiltonian $H$ is not bounded above.
On the other hand, in Minkowski space
the time evolution operators are unitary, and Wightman correlation
functions have a sensible operator interpretation.
We therefore study the correlation functions \Eq{Wightmandefn} in Minkowski space.
Our work differs from most of the existing literature on 4-point functions in Euclidean momentum space~\cite{Maglio:2019grh, Bzowski:2019kwd, Coriano:2019nkw},
as well as other approaches to the conformal bootstrap in momentum space that are based
on the existence of crossing-symmetric correlators~\cite{Polyakov:1973ha, Isono:2018rrb, Isono:2019wex}.

Instead, our work follows the approach of the modern bootstrap program~\cite{Belavin:1984vu, Rattazzi:2008pe}, which makes use of the OPE.
For example, for a 4-point function we can insert a complete set of states to write
\[
\eql{pOPE}
\bra{0} \tilde{\scr{O}}_4(p_4) \cdots \tilde{\scr{O}}_1(p_1) \ket{0}
= \sum_n \bra{0} \tilde{\scr{O}}_4(p_4) \tilde{\scr{O}}_3(p_3) \ket{n}
\bra{n} \tilde{\scr{O}}_2(p_2) \tilde{\scr{O}}_1(p_1) \ket{0}.
\]
The states $\ket{n}$ are naturally chosen to be eigenstates of the translation
generator $P_\mu$ and the quadratic Casimir of the conformal group.
These eigenstates are in one-to-one correspondence to the operators of the theory \cite{Mack:1975je}.
In fact, since $\d_\mu \propto p_\mu$ in momentum space, a complete set of states is given
by $\ket{\tilde{\psi}(p)} \propto \tilde{\psi}(p) \ket{0}$ where $\psi$ is a primary
operator, and the completeness relation can be
written~\cite{Gillioz:2016jnn, Gillioz:2018kwh, Karateev:2018oml}
\[
\eql{pcompleteness}
\id = |0\rangle\langle 0| + \sum_{\psi \, \ne \, \id}
\myint \frac{d^dp}{(2\pi)^d} \Th(p^0)\Th(p^2)\frac{|\tilde\psi(p)\rangle\langle \tilde\psi(p)|}
{\lla \tilde{\psi}(-p) \tilde{\psi}(p)\rra} ,
\]
where
\[
\ket{\tilde{\psi}(p)} = \myint d^d x \ggap e^{-i p \cdot x} \psi(x) \ket{0},
\qquad\qquad
%\bra{\tilde{\psi}(p)} = \ket{\tilde{\psi}(p)}^\dag.
\bra{\tilde{\psi}(p)} = \ket{\tilde{\psi}(p)}^\dag = \bra{0}\tilde{\psi}(-p).
\]
Because \Eq{pOPE} involves only the sum over primary operators, this is already the
conformal block expansion.
In other words, the conformal blocks in momentum space are proportional
to products of 3-point functions of primary operators.%
\footnote{This is true even for operators with spin, as will be discussed in the
main text.}
In this sense, they are conceptually simpler than the
conformal blocks in position space.
In this work,
we will explicitly construct the conformal blocks and study the
OPE in two dimensions.

A remarkable feature of the OPE \eq{pOPE} is that its convergence is apparently independent of the kinematics, since the insertion of a complete set of states is always possible in a Wightman function.
However, this convergence comes with an important caveat: the sum is guaranteed to converge only in the sense of a distribution, and generally does not converge pointwise.
(For a discussion of the convergence of the Wightman OPE from a more mathematical
point of view, see \Ref{Mack:1976pa}.)
Convergence in the sense of a distribution
means that the sum over states is guaranteed to converge only when the correlation
is smeared with a smooth test function in all variables.
An elementary mathematical example that illustrates this is the representation of the
delta function on the interval $-\pi < p \le \pi$ as a Fourier sum:
\[
\eql{deltasum}
2\pi \de(p) = 1 + 2 \sum_{n \, = \, 1}^\infty \cos(n p).
\]
\begin{figure}
\label{fig:delta}
	\centering
	\includegraphics[width=90mm]{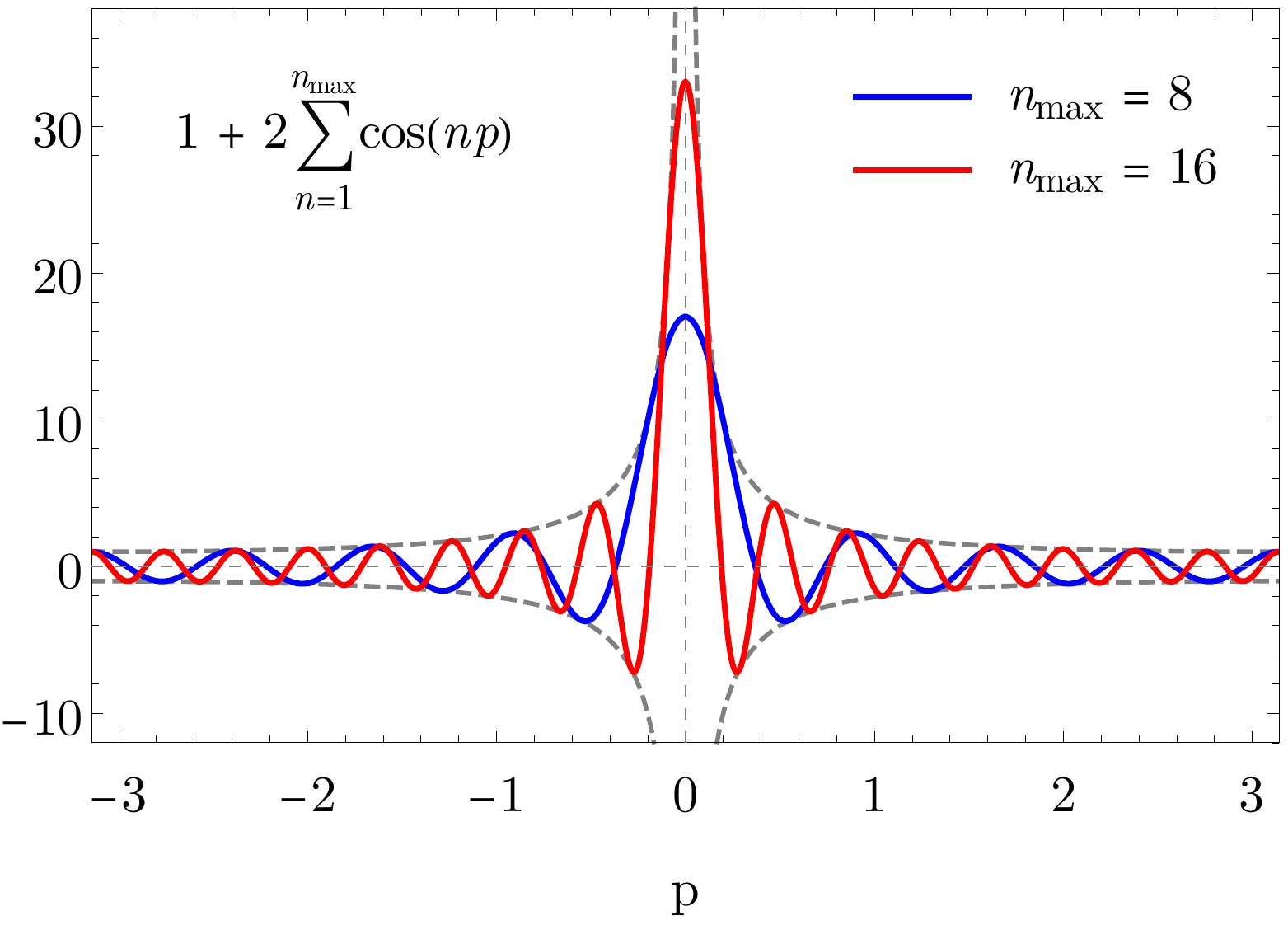}
	\caption{The series defined in \Eq{deltasum} converges to the delta function
	in a distributional sense, but at any given point $p \neq 0$ the partial sums of the series oscillate with an amplitude that is fixed by the value of $p$.
	The envelope of this oscillation is given by $\pm 1/\sin(p/2)$.}
\end{figure}%
We might hope that the sum on the \rhs\ converges to zero for $p \ne 0$,
but this is not the case.
The sum oscillates rapidly with a fixed envelope that is a function of $p$,
as shown in Fig.~\ref{fig:delta}.
However, when integrated against any periodic test function, the sum does converge.
In this work, we focus on CFT in 2 dimensions and show that the momentum-space OPE converges only in this distributional sense for general kinematics.
However, for certain kinematic regions (see Fig.~\ref{fig:diamond}), we show that the momentum space OPE
in fact converges pointwise.
This means that in this kinematic regime we can regard the correlator as an
ordinary function rather than a distribution.
Our proof follows directly from the known convergence properties
of the OPE in Euclidean position space; it does not depend on recent results on
bounding OPE coefficients~\cite{Pappadopulo:2012jk, Rychkov:2015lca, Mukhametzhanov:2018zja}.

We also formulate a bootstrap equation using the OPE \eq{pOPE}.
There are a number of motivations to study the conformal bootstrap in momentum space.
The first is simply that this has not been previously explored.
Also, the momentum space bootstrap is sensitive to correlation functions in a new
kinematic regime where Minkowski space physics is essential.
For example, the contribution of the identity operator can be projected out
by a simple choice of kinematics, and one may hope that the momentum space bootstrap
allows us to obtain more detailed information about higher-dimension operators.
A longer-term hope is to make contact with the study of scattering amplitudes,
which are naturally formulated in momentum space.
For example, the use of the optical theorem in momentum space has been useful in the
study of CFT at large spin~\cite{Komargodski:2012ek},
and in the derivation of a positive sum rule for the $c$ anomaly in
4D CFT~\cite{Gillioz:2018kwh}.

We propose a bootstrap equation based on the microcausality
condition for Wightman functions,
\[
\bra{0} \scr{O}_4(x_4) \bigl[ \scr{O}_3(x_3),\gap \scr{O}_2(x_2)\bigr] \scr{O}_1(x_1) \ket{0}
= 0
\quad \text{for $x_3 - x_2$ spacelike}.
\]
In order to obtain a bootstrap equation in momentum space, we write this as
\[
\eql{fbootstrap}
f(x_3 - x_2)
\bra{0} \scr{O}_4(x_4) \bigl[ \scr{O}_3(x_3),\gap \scr{O}_2(x_2)\bigr] \scr{O}_1(x_1) \ket{0}
= 0,
\]
where $f(x)$ is any function with support only for spacelike $x$.
Taking the Fourier transform of \Eq{fbootstrap} gives a bootstrap equation
for the momentum-space Wightman correlation functions in terms of a convolution
integral over the function $f$.
This equation must be smeared over suitable test functions in order to
obtain a relation among ordinary functions that can be implemented numerically.
(The kinematic region where the OPE converges pointwise does not encompass the
region over which the convolution integrand is nonzero.)
We will not attempt a numerical study of this bootstrap equation in this paper,
but provide all the necessary ingredients in 2D CFT.

This paper is organized as follows.
In section 2, we compute the momentum space conformal blocks for Wightman functions in 2D CFTs.
In section 3, we prove that the conformal block expansion converges pointwise in a
specific kinematic regime.
In section 4, we present the results of calculations in specific CFTs that illustrate
the convergence of the OPE.
In section 5, we formulate the bootstrap equation in momentum space using this OPE.
Section 6 contains our conclusions.
Technical details are given in a number of appendices.

% =============================================================================
\section{Conformal Blocks for Momentum-Space Wightman Functions}
\label{sec:Blocks}
% =============================================================================

The basic objects of our study are the momentum-space Wightman correlation functions defined in \Eq{Wightmandefn}.
(Our conventions are given in Appendix~\ref{appsec:Notations}.)
The OPE for these correlation functions comes from the Hilbert space completeness
relation \Eq{pcompleteness}, where the sum is over all primary operators $\psi$.
Descendant operators are multiples of primary operators in momentum space,
so no sum over descendants is required to define the conformal blocks.
\Eq{pcompleteness} holds as written even for operators with spin in arbitrary
spacetime dimensions,
provided that the basis of intermediate operators is chosen so that
\[
\lla \tilde{\psi}(-p) \tilde{\psi}'(p)\rra = 0
\ \ \text{for}\ \ \tilde{\psi}' \ne \tilde{\psi}.
\]
For example, for a vector operator $\tilde{V}^\mu(p)$ we can choose the independent operators
to be $\tilde{V}^0$ and $\tilde{V}^i$ in the frame where $\vec{p} = 0$.%
\footnote{In general one can use the projection onto spin eigenstates as in Ref.~\cite{Erramilli:2019njx}.}
This implies that the conformal blocks in momentum space are simply products of momentum space correlation functions of primary operators.
For a 4-point function of primary operators, this gives the OPE
\[
\eql{OPEstatesum}
\lla \tilde{\scr{O}}(p_4) \tilde{\scr{O}}(p_3) \tilde{\scr{O}}(p_2) \tilde{\scr{O}}(p_1) \rra
&= \lla \tilde{\scr{O}}(p_4) \tilde{\scr{O}}(p_3) \rra
\lla \tilde{\scr{O}}(p_2) \tilde{\scr{O}}(p_1) \rra (2\pi)^d \de^d(p)
\nn
&\quad{}
+ \sum_{\psi \, \ne \, \id} %\Th(p^0)\Th(p^2)
\frac{\lla \tilde{\scr{O}}(p_4) \tilde{\scr{O}}(p_3) \tilde{\psi}(p) \rra
\lla \tilde{\psi}(-p) \tilde{\scr{O}}(p_2) \tilde{\scr{O}}(p_1) \rra}
{\lla \tilde{\psi}(-p) \tilde{\psi}(p)\rra},
\]
where we define $p = p_1 + p_2 = - p_3 - p_4$.%
\footnote{%
Here and in the following, momentum space correlation functions are understood to
be defined only if the total momentum vanishes, and the momentum flowing
between each pair of operators is in the forward light cone.
For example, for the 4-point function
$\lla \tilde{\scr{O}}(p_4) \tilde{\scr{O}}(p_3) \tilde{\scr{O}}(p_2) \tilde{\scr{O}}(p_1) \rra$
we require that the momenta
$p_1$, $p_1 + p_2$, and $p_1 + p_2 + p_3$ are all in the forward light cone.
(This is known as the spectral condition~\cite{Streater:1989vi}.)
}
The 3-point functions that appear in \Eq{OPEstatesum} contain the dependence on the
OPE coefficients that define the theory.
Note the additional momentum conserving delta function from the identity contribution.
An interesting difference between this OPE and the position space OPE is that the
identity contribution can be distinguished kinematically from the remaining
contributions.

The momentum space conformal blocks have a conceptually simple structure, but in practice the 3-point functions are difficult to compute in general $d$, especially for operators with spin. There are two approaches that can be used. One is a direct calculation of the Fourier transform of the position-space 3-point functions. The other is to solve the conformal Ward identities directly in momentum space, subject to boundary conditions arising from OPE limits. A detailed discussion on computing 3-point functions with the second approach in general spacetime dimension $d$ was given recently in~\cite{Gillioz:2019lgs}. In this paper, we will focus on 2D CFT, where we can simply perform the Fourier transforms for arbitrary operators, including spin.

\subsection{Momentum-Space Conformal Blocks in Two Dimensions}

In 2D Minkowski space we use the standard lightcone coordinates
$x^\pm = x^0 \pm x^1$.
The quantum numbers of an operator are given by the dimension $\De$ and the spin $s$, and we define (as usual)
the conformal weights
$h = \sfrac 12 (\De + s)$,
$\widebar{h} = \sfrac 12 (\De - s)$.
In terms of these the 2- and 3-point Wightman functions of general operators are given in position space by
\[
\bra{0} \O_2(x_2) \O_1(x_1) \ket{0}
&= \left( \frac{e^{-i\pi/2}}{x_{21}^+ - i\ep} \right)^{\!\! 2h}
\left( \frac{e^{-i\pi/2}}{x_{21}^- - i\ep} \right)^{\!\! 2\widebar{h}},
\eql{2ptx}
\\
\bra{0} \O_3(x_3) \O_2(x_2) \O_1(x_1) \ket{0}
&= \la_{123}
\left( \frac{e^{-i\pi/2}}{x_{21}^+ - i\ep} \right)^{\!\! h_{12|3}}
\left( \frac{e^{-i\pi/2}}{x_{31}^+ - i\ep} \right)^{\!\! h_{13|2}}
\left( \frac{e^{-i\pi/2}}{x_{32}^+ - i\ep} \right)^{\!\! h_{23|1}}
\nn
&\qquad{} \times
\left( \frac{e^{-i\pi/2}}{x_{21}^- - i\ep} \right)^{\!\! \widebar{h}_{12|3}}
\left( \frac{e^{-i\pi/2}}{x_{31}^- - i\ep} \right)^{\!\! \widebar{h}_{13|2}}
\left( \frac{e^{-i\pi/2}}{x_{32}^- - i\ep} \right)^{\!\! \widebar{h}_{23|1}},
\eql{3ptx}
\]
where $x_{ab} = x_a - x_b$ and $h_{ab|c} = h_a + h_b - h_c$. The phases and the $i\ep$ prescription in \Eqs{2ptx} and \eq{3ptx} can be understood from the analytic continuation from Euclidean space; this is explained in Appendix~\ref{appsec:Continuation}.

The factorization of the 2- and 3-point functions \Eqs{2ptx} and \eq{3ptx} in position space implies a similar factorization for the momentum space 2- and 3-point functions in terms of light-cone coordinates $p_\pm$  (where $p \cdot x = p_+ x^+ + p_- x^-$).
We write
\[
	\lla \tilde{\O}(-p) \tilde{\O}(p) \rra
	&= D_{2h}(p_+) D_{2\widebar{h}}(p_-),
\eql{2pt}
\\
	\lla \tilde{\O}_3(p_3) \tilde{\O}_2(p_2) \tilde{\O}_1(p_1) \rra
	&= \la_{321} V_{h_3 h_2 h_1}(p_{3+}, p_{2+}, p_{1+})
	V_{\widebar{h}_3 \widebar{h}_2 \widebar{h}_1}(p_{3-}, p_{2-}, p_{1-}).
\eql{3pt}
\]
This implies also a factorization of conformal blocks
\[
\eql{BlockExpansion}
\lla \tilde{\O}_4(p_4) \tilde{\O}_3(p_3) & \tilde{\O}_2(p_2) \tilde{\O}_1(p_1) \rra
=  (2\pi)^d \de^d(p)
\lla \tilde{\O}_4(p_4) \tilde{\O}_3(p_3) \rra
\lla \tilde{\O}_2(p_2) \tilde{\O}_1(p_1) \rra
\nn
&{}\qquad
+ \sum_{\psi \, \ne \, \id} \lambda_{\phi\phi\psi}^2 W_{h_\psi}(p_{4+}, p_{3+}, p_{2+}, p_{1+})
W_{\widebar{h}_\psi}(p_{4-}, p_{3-}, p_{2-}, p_{1-}),
\]
We will refer to the objects $W_h(k_4, k_3, k_2, k_1)$ as \emph{holomorphic conformal blocks}.
Note that \Eq{BlockExpansion} defines the conformal blocks only for the global conformal symmetry, not the
Virasoro blocks of \Ref{Belavin:1984vu}.
We will generally use the letter $k$ to denote the lightcone
components of 2D momenta $p_\pm$ in the following.%
\footnote{In our conventions, the condition that the momentum
$p$ is in the forward light cone
is $p_\pm \ge 0$.
We therefore understand that holomorphic quantities
such as $W_h(k_4, k_3, k_2, k_1)$ are nonzero
only if $k_1 + \cdots + k_4 = 0$,
and $k_1$, $k_1 + k_2$, $k_1 + k_2 + k_3 > 0$.}

To evaluate the Fourier transforms, we use the identity
\[
\eql{propFT}
\int\limits_{-\infty}^\infty dx\ggap e^{ikx} \ggap
\left( \frac{e^{-i\pi/2}}{x - i\ep} \right)^{\!\! \al}
= \Th(k) \frac{2\pi}{\Ga(\al)} k^{\al - 1}.
\]
The integral is convergent for $\Re\al > 0$, but the \rhs\ provides an analytic
continuation of the integral for general complex $\al$.
Analytic continuation of this kind is justified by the fact that the
functions we are computing are uniquely determined by the conformal Ward identities.
(up to the OPE coefficients). We have
\[
\lla \tilde{\O}(-p) \tilde{\O}(p) \rra
= \myint \sfrac 12 dx^+ dx^- \ggap
e^{i (p_+ x^+ + p_- x^-)}
\left( \frac{e^{-i\pi/2}}{x^+ - i\ep} \right)^{\!\! 2h}
\left( \frac{e^{-i\pi/2}}{x^- - i\ep} \right)^{\!\! 2\widebar{h}},
\]
which gives
\[
\eql{V2}
D_{2h}(k)
= \Th(k)\frac{2\pi}{\sqrt{2}\ggap \Ga(2h)} k^{2h-1}.
\]
To compute the 3-point function, we can view the factors
\[
\eql{prop}
\left( \frac{e^{-i\pi/2}}{x_{12} - i\ep} \right)^{2h}
= \includegraphics[valign=c]{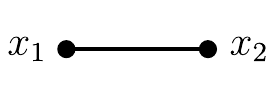}
= \int\limits_{-\infty}^\infty \frac{dk}{2\pi} e^{-ikx_{12}} \sqrt{2}D_{2h}(k) ,
\]
as propagators in position space, so that \Eq{propFT} gives the momentum space propagator. In this way, we obtain
\[
V_{h_3 h_2 h_1}& (k_3, k_2, k_1)
= \includegraphics[width=60mm, valign=c]{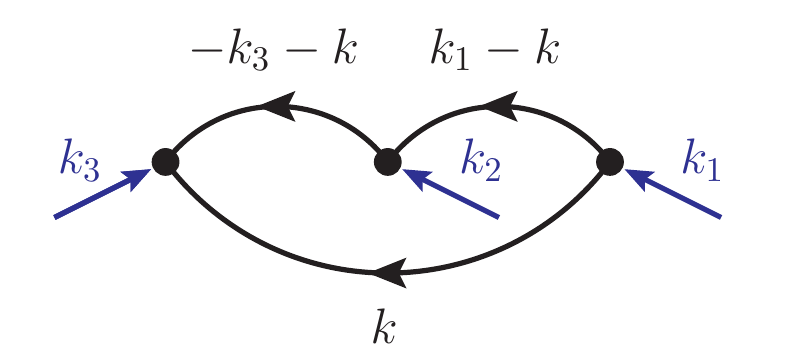}
\nn
&=\sqrt{2} \int\limits_{-\infty}^\infty \frac{dk}{2\pi} D_{h_{12|3}}\left(k_1-k\right) D_{h_{13|2}}\left(k\right) D_{h_{23|1}}\left(-k_3-k\right)
\nn
&= \Th\left(-k_3\right)\Th\left(k_1\right) \frac{1}{2}(2\pi)^2
\nn
&\quad\times \left[ \begin{array}{l}
\Th\left( k_2 \right)\dfrac{{{{\left( -{k_3} \right)}^{{h_{23|1}} - 1}}{{\left( {{k_1}} \right)}^{2{h_1} - 1}}}}{{\Gamma \left( {{h_{23|1}}} \right)\Gamma \left( 2h_1 \right)}}{}_2{F_1}\!\left( {1 - {h_{23|1}},{h_{13|2}};2h_1;-\dfrac{k_1}{{ k_3}}} \right)\\
{}+ \Th\left( -k_2 \right)\dfrac{{{{\left( -k_3 \right)}^{2h_3 - 1}}{{\left( k_1 \right)}^{{h_{12|3}} - 1}}}}{{\Gamma \left( {{h_{12|3}}} \right)\Gamma \left( {2h_3} \right)}}{}_2{F_1}\!\left( {1 - {h_{12|3}},{h_{13|2}};2h_3;-\dfrac{k_3}{k_1}} \right)
\end{array} \right] .
\eql{V3}
\]
Note that the above calculation is performed with the aid of analytic continuation. Certain intermediate steps hold only for the regime $h_{ij|k}<0$, but the final result can be analytically continued
to all physical values of the $h_i$. 
These 3-point function agree with the recent results of \Ref{Anand:2019lkt}.
The holomorphic conformal block is then given by
\[
	W_h(k_4, k_3, k_2, k_1)
	= \frac{V_{h_4 h_3 h}(k_4, k_3, k) V_{h h_2 h_1}(-k, k_2, k_1)}{D_{2h}(k)},
\eql{HolomorphicConformalBlocks}
\]
where $k = k_1 + k_2 = - k_3 - k_4$.

These conformal blocks are analytic functions of the conformal weights as long as $h > 0$.
In the special case where one of the conformal weights is zero, they vanishes for generic kinematics.%
\footnote{When one of the external weights $h_1$ to $h_4$ vanishes, the corresponding operator is holomorphic and the 4-point function can in general be simplified. An example of correlation function with 4 holomorphic operators is given later in Section~\ref{sec:examples}. If instead it is the conformal weight of the exchanged operator that is zero, then $W_0(k_4,k_3,k_2,k_1) \propto \delta(k_2 + k_1)$. This is the case for instance if the exchanged operator is a Virasoro descendant of the identity.
}
The conformal blocks also have zeroes when the conformal weights obey special relations, namely when some of the $h_{ij|k}$ are negative integers. The simplest situation in which this happens is generalized free field theory, where $h_\psi = h_1 + h_2 + n = h_3 + h_4 + n$ with $n \in \mathbb{N}$. In this case the block is identically zero whenever $p_{2\pm} < 0$ or $p_{3\pm} < 0$, i.e.~when either $p_2$ or $p_3$ (or both) do not lie in the forward light cone.
This will be relevant in Section~\ref{sec:convergence:GFF}.

% =============================================================================
\section{Pointwise Convergence of the OPE}
\label{sec:Convergence}
% =============================================================================

\begin{figure}
	\includegraphics[width=0.47\linewidth]{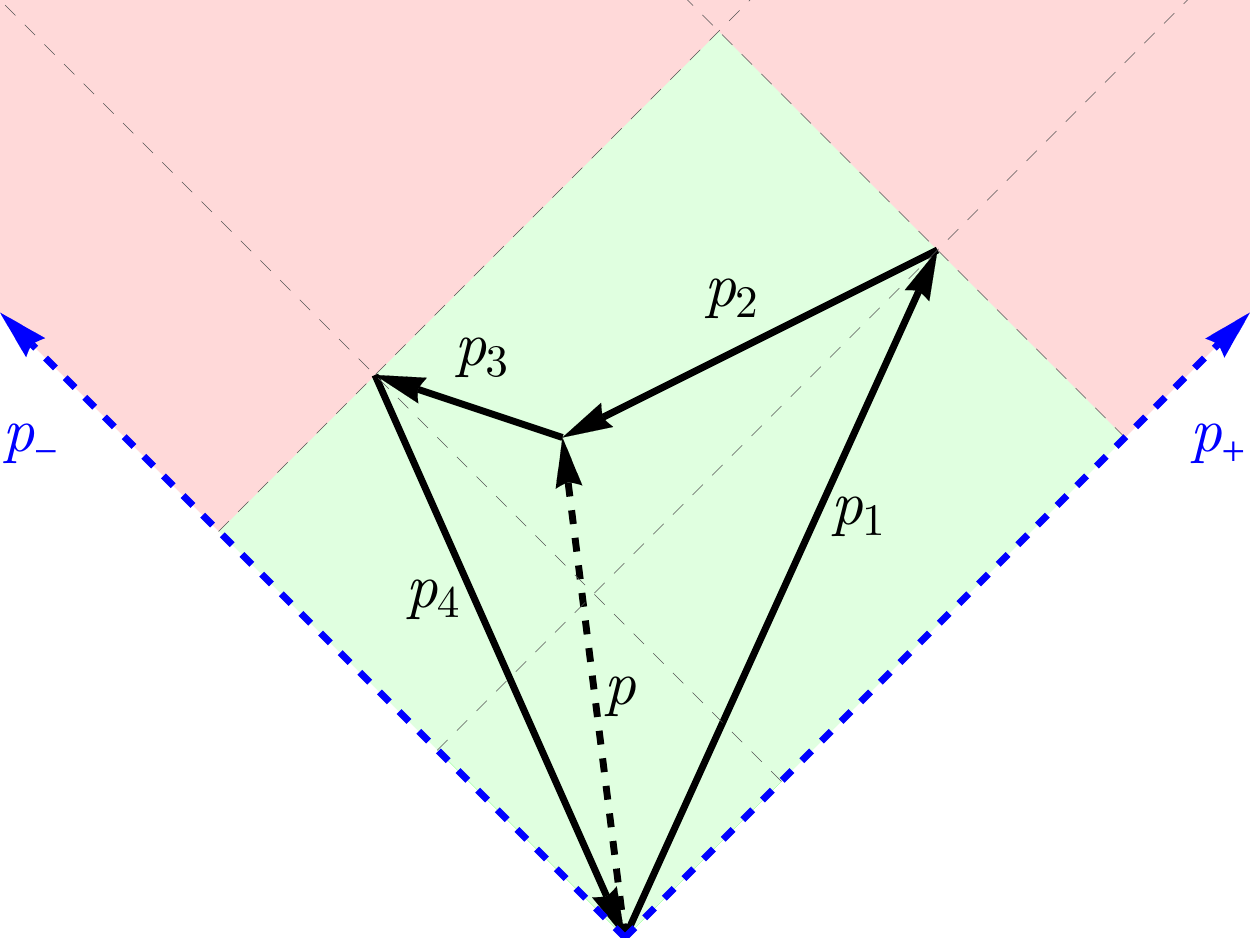}
	\hfill
	\includegraphics[width=0.47\linewidth]{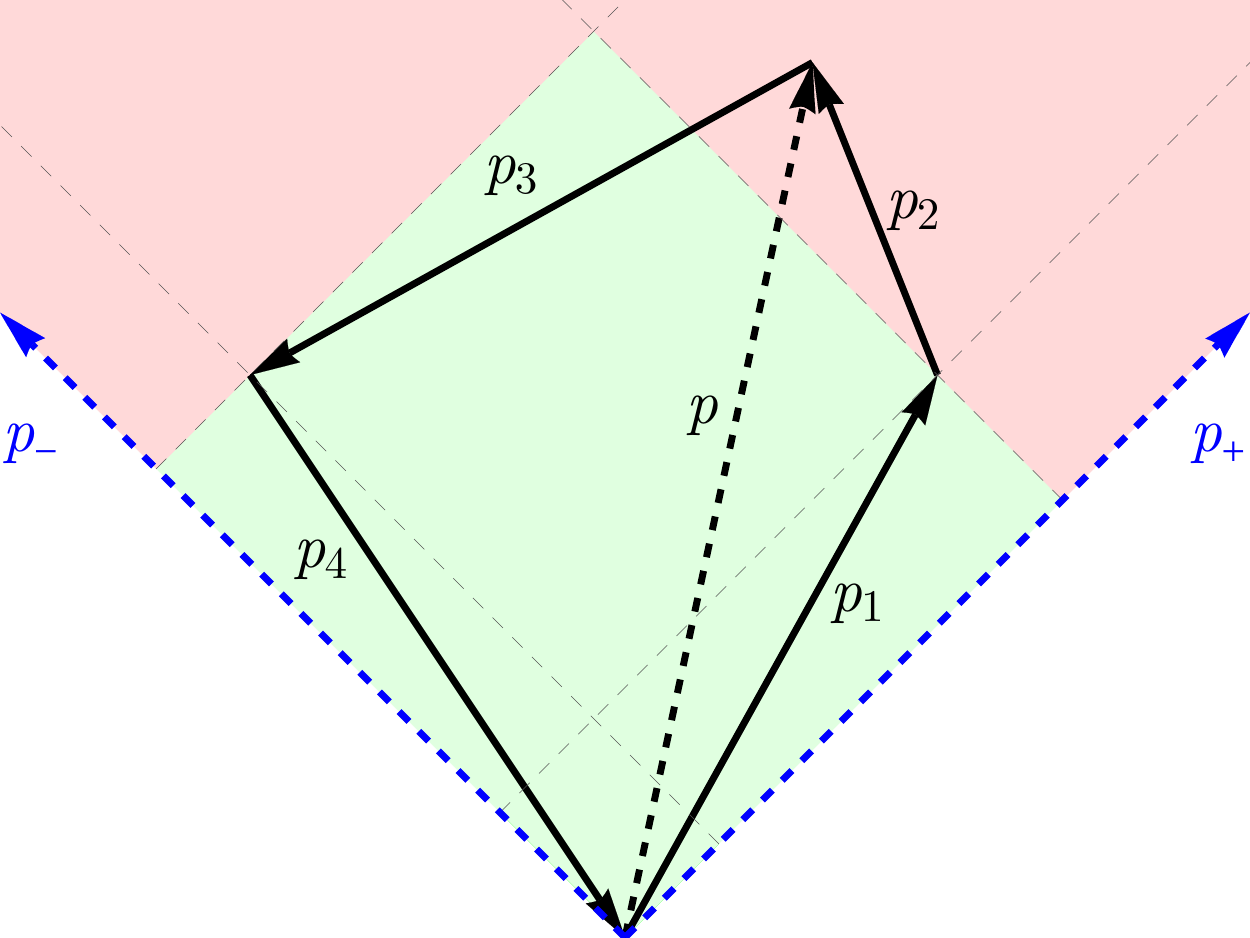}
	\caption{Two possible configurations of momenta.
	The Wightman 4-point function is non-zero only if all three momenta $p_1$, $p$ and $-p_4$
	lie in the forward light cone (shaded region), which is the case here.
	The criterion for pointwise OPE convergence is that $p$ lies in the diamond delimited by $p_1$ and $-p_4$
	(green region). The OPE is therefore pointwise convergent for the configuration in the left panel,
	but not for the configuration in the right panel.}
\label{fig:diamond}
\end{figure}

The Wightman 4-point function is non-zero only if all three momenta $p_1$, $p$ and $-p_4$ lie in the forward light cone, shown as the shaded regions in Fig.~\ref{fig:diamond}.
Our expansion derived by inserting a complete set of momentum eigenstates is
guaranteed to converge as a distribution everywhere in this region.
In this section, we prove that in the subregion $p_\pm < \max\left(p_{1\pm}, -p_{4\pm}\right)$
(the green region in Fig.~\ref{fig:diamond})
the momentum-space conformal block expansion is in fact pointwise convergent.
Therefore, in this kinematic regime the correlator defines an ordinary
function rather than a distribution.
The convergence follows from the fact that the partial sums of this series expansion are bounded by the partial sums of the Euclidean position-space conformal block expansion, which we know is pointwise convergent.
Concrete examples that show the absence of pointwise convergence outside this region will be presented in Section~\ref{sec:examples}.

The convergence of the sum \eq{pOPE}
is determined by the asymptotics of the terms for large
$h$ and/or $\widebar{h}$.
This can be obtained directly from the results of Section~\ref{sec:Blocks}.
Applying the formulas of appendix~\ref{appsec:Asymptotics} to the definition \eqref{eq:HolomorphicConformalBlocks}, we find%
\footnote{The notation $A \stackrel{h \to \infty}{\simeq} B$ in this and later equations means $\lim\limits_{h \to \infty} A/B = 1$.}
\[
	W_h(k_4, k_3, k_2, k_1) & \stackrel{h \to \infty}{\simeq} (2\pi)^{3/2}
	\left( \prod_{i = 1}^4 |k_i|^{h_i - 3/4} \right)
	\frac{2^{2h-1}}{h^{h_1 + h_2 + h_3 + h_4 - 3/2}}
	\mathcal{S}_{12}(k_1,k_2) \mathcal{S}_{43}(-k_4,-k_3),
\eql{Wasymptotics}
\]
where 
\[
	\mathcal{S}_{12}(k_1,k_2)
	= \left\{ \begin{array}{ll}
		\sin\left[ \pi (h_1 + h_2 - h) \right]
		\left( \dfrac{\sqrt{k_1 + k_2}}{\sqrt{k_1} + \sqrt{-k_2}} \right)^{2h-1}
		& \text{if}~k_2 < 0,
		\\[4mm]
		\sin\left[ \left( h - \frac{1}{2} \right) \arccos\left( \dfrac{k_2 - k_1}{k_1 + k_2} \right)
		- \pi \left( h_1 - \frac{3}{4} \right) \right]
		\quad
		& \text{if}~k_2 \geq 0,
	\end{array} \right.
\eql{S}
\]
with a similar expression being valid for $\mathcal{S}_{43}$.
Note that $|\mathcal{S}_{12}| \leq 1$, with an important difference between the cases $k_2 < 0$ and $k_2 \geq 0$: in the latter $\mathcal{S}_{12}$ oscillates in the interval $(-1, 1)$ when $h$ increases, while in the former case $\mathcal{S}_{12}$ is exponentially suppressed.

This asymptotic behavior can be compared with that of the Euclidean position space
holomorphic conformal block~\eqref{eqn:PSBlock}, for which
\[
	G_h(\eta) \stackrel{h \to \infty}{\simeq}
	\sqrt{\eta} (1-\eta)^{(h_4 - h_3 - h_2 + h_1 - 1/2)/2} 2^{2h-1} \left( \frac{\sqrt{\eta}}{1 + \sqrt{1-\eta}} \right)^{2h-1}
\eql{Gasymptotics}
\]
in terms of the cross-ratio $\eta$. This expression shares with \eqref{eq:Wasymptotics} the exponential growth factor $2^{2h}$. Moreover, the last term resembles the suppression factor present in $\mathcal{S}_{12}$ when $k_2 < 0$ with the identification $\eta = (k_1 + k_2)/k_1$.
In fact, when $k_2 < 0$ or $k_3 > 0$ (or both), it is always possible to choose a point $\eta_*$ in the interval
\[
	\min\left( \frac{k_1 + k_2}{k_1}, \frac{k_3 + k_4}{k_4} \right) < \eta_* < 1
\]
so that
\[
	\mathop {\lim }\limits_{h \to \infty }
	\frac{W_h\left(k_4, k_3, k_2, k_1\right)}{G_h(\eta_*)} = 0.
\eql{RatioLimit}
\]
This choice of $\eta_*$ is such that $W_h$ decays exponentially faster than $G_h$ as $h \to \infty$. For this reason, the presence of numerical factors or powers of $h$ in Eqs.~\eqref{eq:Wasymptotics} and \eqref{eq:Gasymptotics} does not affect the result, which is independent of the scaling dimension and spin of the external operators.
Since for generic kinematics the holomorphic conformal block $W_h$ remains finite for all values of $h$, and since $G_h$ is real and positive over the interval $0 < \eta < 1$, we have moreover
\[
	\left| \frac{W_h\left(k_4, k_3, k_2, k_1\right)}{G_h(\eta_*)} \right| < \infty
	\quad
	\text{for all}~h \geq 0.
\eql{RatioFiniteness}
\]
It is well known that the conformal block expansion in Euclidean position space is absolutely convergent for any $\eta = \mathbb{C} \setminus (1, \infty)$~\cite{Hogervorst:2013sma}, and in particular on the real interval $0<\eta<1$.
Thus, \Eqs{RatioLimit} and \eq{RatioFiniteness} together prove that the momentum-space OPE is pointwise convergent as long as $k_2 < 0$ or $k_3 > 0$ (or both), corresponding to the green region in Fig.~\ref{fig:diamond}.

% =============================================================================
\section{Examples}
\label{sec:examples}
% =============================================================================

In this section we provide three examples in which all the OPE data is known and the
convergence of the momentum space conformal block expansion can be studied
explicitly.
These examples show that the OPE in general converges only as a distribution
outside the region for which we proved pointwise convergence.

\subsection{Generalized Free Field Theory}
\label{sec:convergence:GFF}

We consider first the simplest CFT: generalized free field theory.
By definition, the 4-point function of a generalized free scalar field $\phi$ with scaling dimensions $\Delta_\phi$ obeys
\[
\begin{split}
	\lla \tilde{\phi}(p_4) \tilde{\phi}(p_3) \tilde{\phi}(p_2) \tilde{\phi}(p_1) \rra
	&= (2\pi)^2\delta^2(p_1+p_2) \lla \tilde{\phi}(p_4) \tilde{\phi}(p_3) \rra
	\lla \tilde{\phi}(p_2) \tilde{\phi}(p_1) \rra
	\\
	& \qquad{}
	+ (2\pi)^2\delta^2(p_1+p_3) \lla \tilde{\phi}(p_4) \tilde{\phi}(p_2) \rra
	\lla \tilde{\phi}(p_3) \tilde{\phi}(p_1) \rra
	\\
	& \qquad{}
	+ (2\pi)^2\delta^2(p_1+p_4) \lla \tilde{\phi}(p_4) \tilde{\phi}(p_1) \rra
	\lla \tilde{\phi}(p_3) \tilde{\phi}(p_2) \rra.
\end{split}
\]
Using Eq.~\eqref{eq:2pt}, this amounts to
\[
\begin{split}
	\lla \tilde{\phi}(p_4) \tilde{\phi}(p_3) \tilde{\phi}(p_2) \tilde{\phi}(p_1) \rra
	&= \frac{(2\pi)^6}{2^{4\Delta_\phi-2} \left[ \Gamma(\Delta_\phi) \right]^4}
	(p_4^2 p_3^2 p_2^2 p_1^2)^{(\Delta_\phi - 1)/2}
	\\
	& \quad{} \times
	\left[ \delta^2(p_1 + p_2) + \delta^2(p_1 + p_3)
	+ \delta^2(p_1 + p_4) \Theta(p_{2+}) \Theta(p_{2-}) \right].
\eql{4pt:GFF}
\end{split}
\]
Of the three delta functions on the \rhs, the first one arises from the identity term in the OPE.
We now show that the other two are reproduced by the conformal block expansion as a distribution.

In generalized free field theory, all the operators that enter the OPE $\phi \times \phi$ are schematically of the form $\phi \partial_+^m \partial_-^{\widebar{m}} \phi$. Their conformal weights obey $(h, \widebar{h}) = (\Delta_\phi + m, \Delta_\phi + \widebar{m})$ for some $m, \widebar{m} \in \mathbb{N}$, and the OPE coefficients
are given in \Ref{Fitzpatrick:2011dm}.
We can therefore study the convergence of the OPE using the known CFT data.
The contribution of an operator with large conformal weights is given by
\[
\begin{split}
	\lambda_{m, \widebar{m}}^2
	W_h(p_{i+}) &
	W_{\widebar{h}}(p_{i-})
	\\{}
	\stackrel{m, \widebar{m} \to \infty}{\simeq} &
	\left[ {1 + {{\left( { - 1} \right)}^{m + \bar m}}} \right]
	\frac{(2\pi)^4}{2^{4\Delta_\phi - 7} \Gamma(\Delta_\phi)^4}
	(p_4^2 p_3^2 p_2^2 p_1^2)^{(2\Delta_\phi - 3)/4}
	\mathcal{S}_{12} \mathcal{S}_{43}
	\widebar{\mathcal{S}}_{12} \widebar{\mathcal{S}}_{43}.
\end{split}
\]
Here $\mathcal{S}_{ab}$ is defined in Eq.~\eqref{eq:S} in terms of $p_{i+}$ and $h$, while $\widebar{\mathcal{S}}_{ab}$ is its equivalent for the anti-holomorphic part depending on $p_{i-}$ and $\widebar{h}$.
It turns out that the $\mathcal{S}_{ab}$ are identically zero in generalized free field theory whenever $p_{2\pm} < 0$ or $p_{3\pm} > 0$, which is consistent with the vanishing of Eq.~\eqref{eq:4pt:GFF} in the same kinematic range.
On the other hand, when both $p_2$ and $p_3$ lie inside the forward light cone, each of the $\mathcal{S}_{ab}$ is a phase and all the operator give a contribution of similar magnitude.
It is clear in this case that the OPE is not pointwise convergent.
In Appendix~\ref{appsec:GFF} we show that it in fact converges in the
distributional sense to give the delta functions in \Eq{4pt:GFF}.

\subsection{Ising Model}

\begin{figure}
	\includegraphics[width=0.49\linewidth]{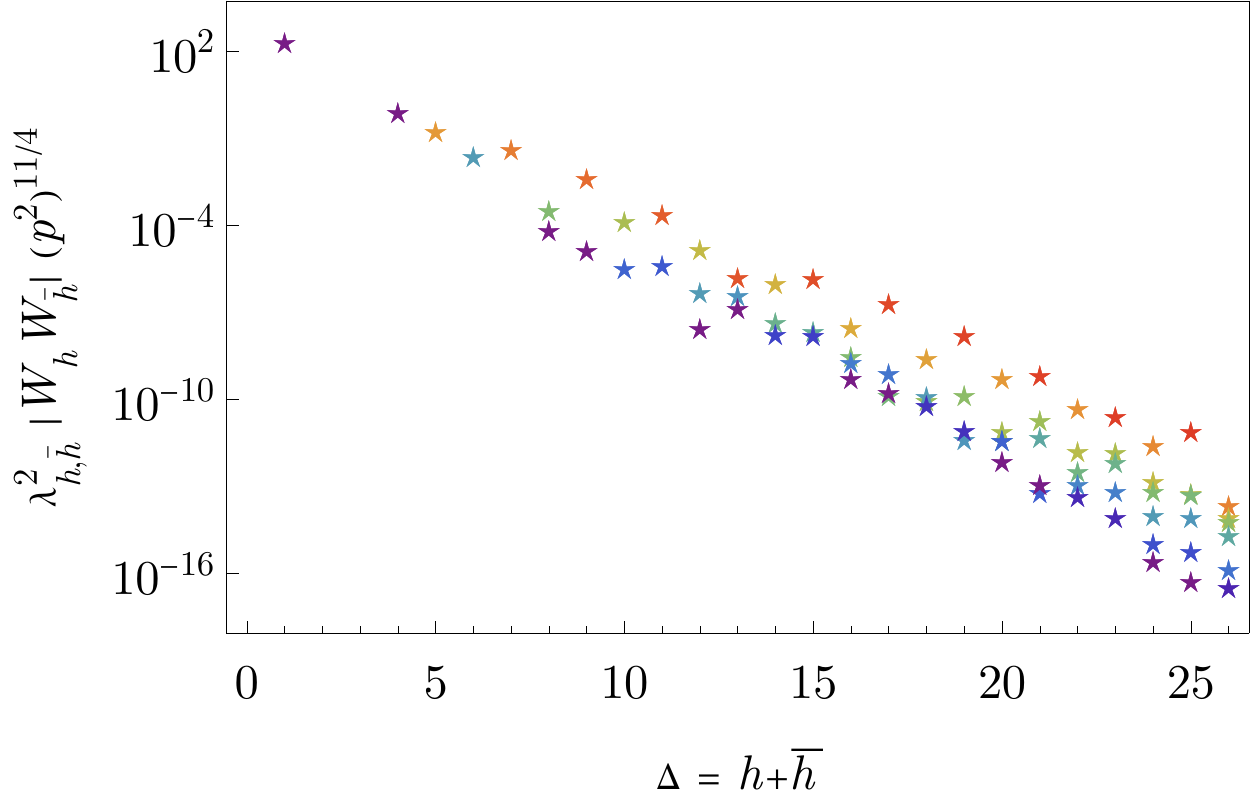}
	\hfill
	\includegraphics[width=0.49\linewidth]{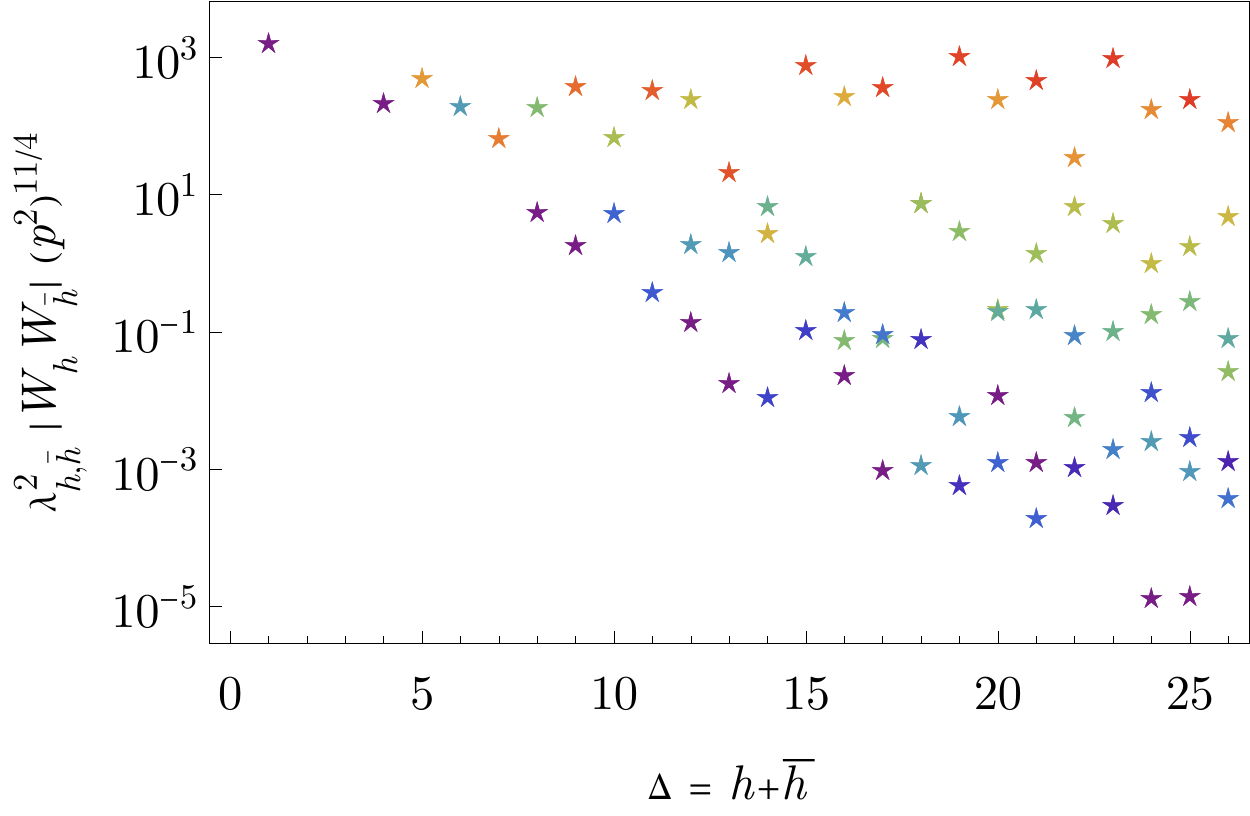}
	\caption{Contribution to the conformal block expansion of the 4-point function
	$\langle\sigma\sigma\sigma\sigma\rangle$ in the Ising model, for two different
	configurations of momenta corresponding to those of Fig.~\ref{fig:diamond}.
	Only the absolute value of the contributions is shown, not their sign.
	The horizontal axis indicates the scaling dimension of the operators
	and the color encodes the spin (from violet for spin $s = 0$ to red for spin $s = \Delta$).
	In the convergent case (left panel) the contributions decrease exponentially
	with the scaling dimension, while in the divergent case (right panel) operators of
	all scaling dimensions give contributions of similar size.}
\label{fig:Ising:OPEconvergence}
\end{figure}

The Ising model is the simplest minimal model for which all correlation functions are known in position space. There are two Virasoro primary operators $\sigma$ and $\varepsilon$ with conformal weights $(h, \widebar{h}) = \left( \frac{1}{16}, \frac{1}{16} \right)$ and $\left( \frac{1}{2}, \frac{1}{2} \right)$ respectively.
Focusing on the 4-point function of the operator $\sigma$, we have, in the notation of \Eq{4ptxapp}~\cite{Belavin:1984vu, Caron-Huot:2017vep},
\[
\begin{split}
	G_{\langle\sigma\sigma\sigma\sigma\rangle}(\eta,\widebar{\eta}) = \frac{1}{(1-\eta)^{1/8}(1-\widebar{\eta})^{1/8}}
	\bigg[ & \left( \frac{1 + \sqrt{1 - \eta}}{2} \right)^{1/2}
	\left( \frac{1 + \sqrt{1 - \widebar{\eta}}}{2} \right)^{1/2}
	\\
	& + \left( \frac{1 - \sqrt{1 - \eta}}{2} \right)^{1/2}
	\left( \frac{1 - \sqrt{1 - \widebar{\eta}}}{2} \right)^{1/2} \bigg].
\eql{Gsigma}
\end{split}
\]
The conformal weights and OPE coefficients of all the operators entering the OPE $\sigma \times \sigma$ can be extracted from the expansion of $G$ around $(\eta, \widebar{\eta}) = (0,0)$. Using this data together with our momentum-space conformal blocks allows to study the convergence of the expansion.
Fig.~\ref{fig:Ising:OPEconvergence} shows the result of such an analysis for the first 100 operators in the OPE with the two configurations of momenta of Fig.~\ref{fig:diamond}.
Considering more configurations of momenta, we are able to verify empirically that the momentum-space OPE converges in the diamond-like region $p_\pm < \max(p_{i\pm}, p_{f\pm})$ and diverges otherwise.

This example is important because it shows that the OPE is not convergent pointwise
in configurations such as the \rhs\ of Fig.~\ref{fig:diamond},
where the generalized free field theory example is trivial since all conformal blocks vanish.
It also shows that the absence of pointwise convergence is not necessarily associated
with disconnected contributions that give rise to delta functions in momentum space.
For example, the double commutator
$\bra{0} [\phi_4, \phi_3] [\phi_2, \phi_1] \ket{0}$
has no disconnected contributions,
and yet when both $p_3$ and $p_2$ are spacelike it is equal to the Wightman function,
and it has an OPE that does not converge pointwise.

\subsection{Energy-Momentum Tensor}

The energy-momentum tensor is an operator present in any local CFT. It has two components $T$ and $\widebar{T}$ with conformal weights $(h, \widebar{h}) = (2,0)$ and $(0,2)$ respectively.
$T$ and $\widebar{T}$ are Virasoro descendants of the vacuum, and therefore their correlation functions are completely fixed in terms of the central charge $c$.
In Euclidean position space, the 2- and 4-point functions of $T$ are given by%
\footnote{We do not use the convention~\eqref{eq:2ptapp} because $T$ is a conserved current and its normalization is fixed by a Ward identity.}
\[
	\langle T(x^+_2) T(x^+_1) \rangle
	&= \frac{c/2}{(x^+_{21})^4},
	\\
	\!\!
	\langle T(x^+_4) T(x^+_3) T(x^+_2) T(x^+_1) \rangle
	&= \frac{c^2/4}{(x^+_{43})^4 (x^+_{21})^4}
	+ \frac{c}{(x^+_{42})^2 (x^+_{41})^2 (x^+_{32})^2 (x^+_{31})^2}
	+ \text{permutations}.
\eql{TTTT:positionspace}
\]
In the 4-point function, the term quadratic in $c$ corresponds to a generalized free field theory correlator and will be referred to as the `disconnected' part of the correlator, while the term linear in $c$ is the `connected' part.
This partition corresponds to the distinction between disconnected and connected Feynman diagrams in the free fermion ($c = \frac{1}{2}$) and the free boson ($c = 1$) theories.

\begin{figure}
	\includegraphics[width=0.49\linewidth]{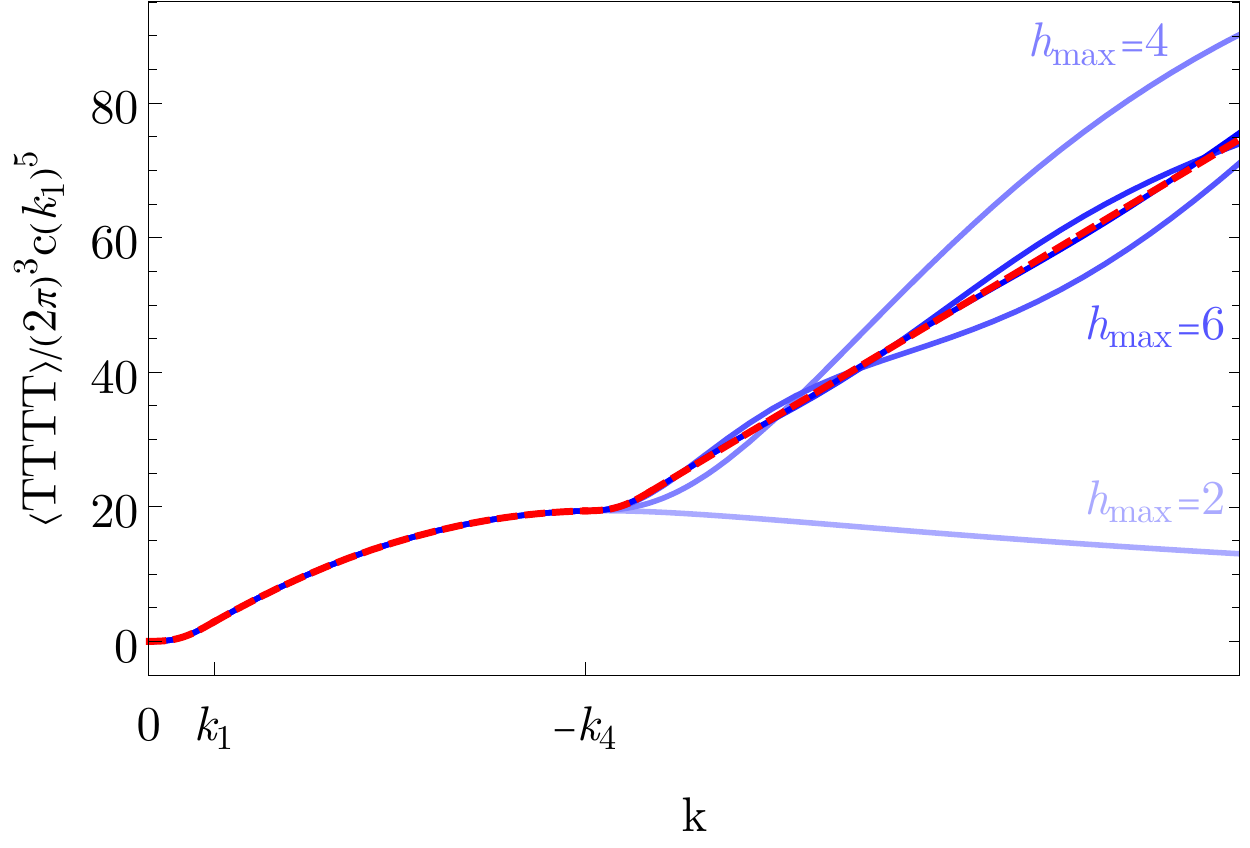}
	\hfill
	\includegraphics[width=0.49\linewidth]{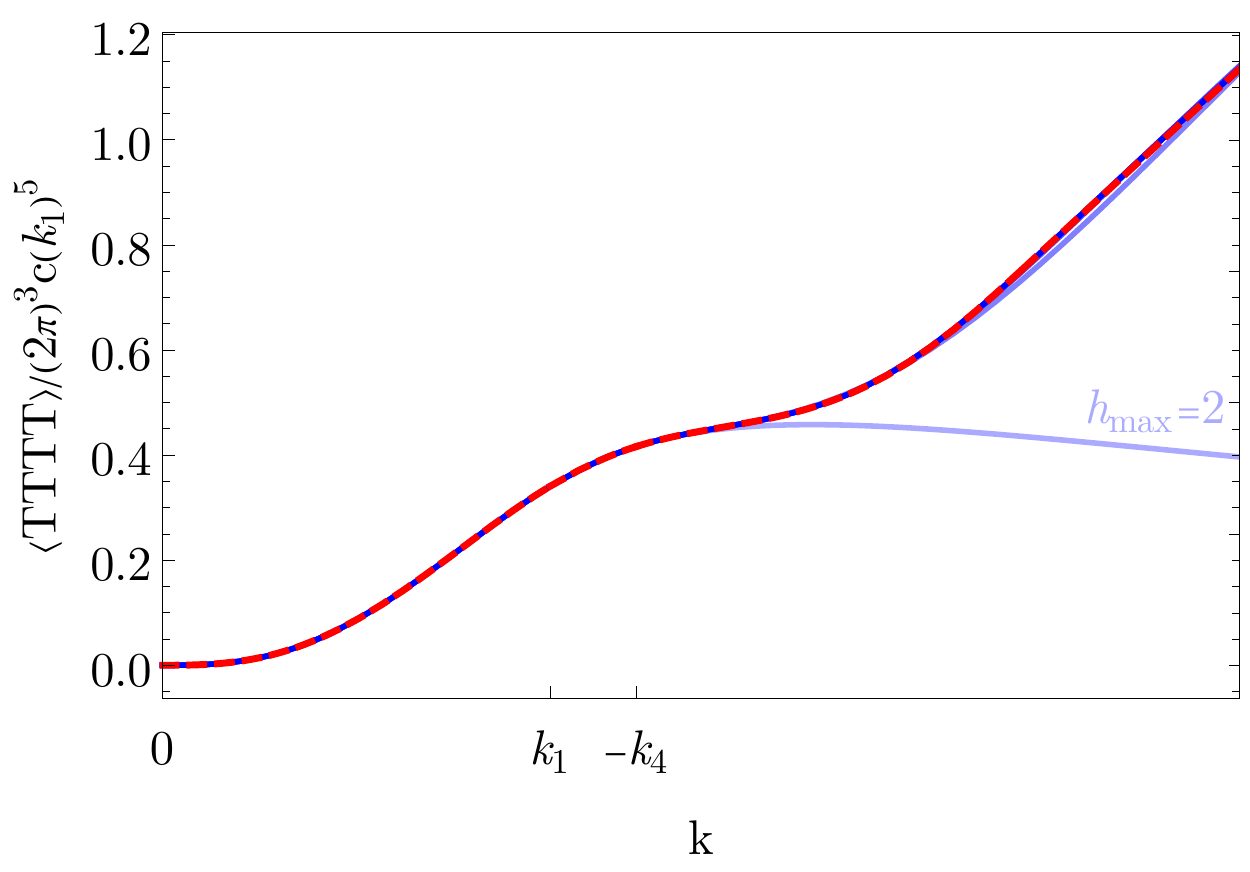}
	\caption{The connected part of the 4-point correlation function $\langle TTTT \rangle$
	given by \Eq{TTTT:W} as a function of the momentum $k = k_1 + k_2$ (red dashed line),
	compared with different truncations of the conformal block expansion
	(blue lines, with $h_\text{max} = 2, 4, 6, 8, 10$ from lightest to darkest).
	Two different kinematic configurations are shown, one in which $k_1 \ll -k_4$ (left panel)
	and the other in which $k_1 \simeq -k_4$ (right panel).
	}
\label{fig:TTTT}
\end{figure}

Correspondingly, the OPE coefficient associated to an intermediate operator with conformal weight $(h, 0)$ in the $T \times T$ OPE obeys
\[
\begin{split}
	\lambda_h^2 &= \frac{c^2}{12}
	\left[ \frac{(2h - 1) (h-3) (h-2) (h-1) h! (h+2)!}{3 (2h)!} - \delta_{h,0} \right]
	\\
	& \qquad{}
	+ 2c \, \frac{(h^2 - h - 1) (h-2)! (h-1)!}{(2h - 3)!},
\eql{TTTT:lambda}
\end{split}
\]
where $h$ can take any even integer value.
The disconnected part of this OPE coefficient (the term proportional to $c^2$)
is the generalized free field theory expression, and we know that it leads to a momentum-space OPE that does not converge in a pointwise manner.
On the other hand, the connected part (term in $c$) decays faster than the disconnected part at large $h$,
and therefore its OPE is expected to be convergent.
Indeed, we find that the Fourier transform of the connected part of Eq.~\eqref{eq:TTTT:positionspace} does not involve delta functions but is piecewise polynomial:
\[
	\lla T(k_4) T(k_3) & T(k_2) T(k_1) \rra_\text{conn}
	\nn
	&=
	\frac{(2\pi)^3 c}{3\sqrt{2}}
	\left\{
	\begin{array}{ll}
		k^3 (k_1 - k_2) (k_3 - k_4) &
		\text{if} ~ 0 \leq k < k_1,
		\\[3mm]
		k_1^3 (k_3 - k_4) (k_2 - k_3 - k_4)&
		\text{if} ~ k_1 \leq k < -k_4,
		\\[3mm]
		k_1^3 (k_2 - k_3) (k_2 + k_3 - k_4) &
		\multirow{2}{*}{$\text{if} ~ -k_4 \leq k < k_1 - k_4,$}
		\\
		- (k_1 - k_3) (k_2 - k_4) (k_2 + k_4)^3 &
		\\[3mm]
		k_1^3 (k_2 - k_3) (k_2 + k_3 - k_4) &
		\text{if} ~ k \geq k_1 - k_4,
	\end{array} \right.
\eql{TTTT:W}
\]
where we have assumed for simplicity of notation that $k_1 < -k_4$ and split the different cases according to the value of $k \equiv k_1 + k_2 = - k_3 -k_4$.
This function of $k$ is continuous everywhere but not differentiable at $k = k_1$, $-k_4$ and $k_1 - k_4$.
It can be compared with the conformal block expansion using the OPE coefficients \eqref{eq:TTTT:lambda}. In the kinematic range $0 < k < -k_4$, we find that the sum is saturated by the first term, i.e.~the contribution of $T$ itself, while the contributions of all other operators vanish.
In the range $k > -k_4$, on the contrary, all operators contribute to the expansion.%
\footnote{This had to be the case since the conformal blocks are analytic over the entire range $k \in [-k_4, \infty)$ whereas the full correlation function \eqref{eq:TTTT:W} is not differentiable at the point $k = k_1 - k_4$.
}
The expansion is found to be convergent over the full kinematic range, although the rate of convergence varies depending on the ratios $k_1/k$ and $k_4/k$. Fig.~\ref{fig:TTTT} illustrates this convergence in two particular cases.

% =============================================================================
\section{Momentum-Space Bootstrap}
\label{sec:Bootstrap}
% =============================================================================

In this section we formulate the bootstrap equations for Wightman functions
in momentum space.
We begin by considering Wightman 4-point functions in position space.
We can write bootstrap equations for these correlation functions by
using the microcausality condition, namely that local operators commute at
spacelike separation. Considering identical scalars for simplicity, this is
\[
\eql{Wightbootposn}
\bra{0} \phi(x_4) \bigl[ \phi(x_3),\gap \phi(x_2)\bigr] \phi(x_1) \ket{0} = 0
\qquad \text{for}~x_3 - x_2~\text{spacelike}.
\]
Inserting a complete set of states in both terms in the commutator gives a
crossing equation.

We obviously cannot simply Fourier transform this equation
because the Fourier transform integrates over values of the position where
$x_3 - x_2$ is timelike.
But we can write
\[
\eql{CommutatorMatrixElement}
\bra{0} \phi(x_4) \bigl[ \phi(x_3), \phi(x_2)\bigr] \phi(x_1) \ket{0} f(x_3 - x_2) = 0,
\]
where $f(x)$ has support only for spacelike $x$.
This equation is valid for all positions, and we can obtain a
bootstrap equation in momentum space by Fourier transforming it.
One simple choice for $f$ is
\[
\eql{fchoice}
f(x) = \de(x^0).
\]
To write the bootstrap equation explicitly,
it is convenient to introduce the momentum variable
$q = \frac{1}{2} \left( p_2 - p_3 \right)$,
and define
\[
	W(p_4, p_1 | q) = \lla \tilde{\phi}(p_4)
	\tilde{\phi}\left( - q - \tfrac{p_4 + p_1}{2} \right)
	\tilde{\phi}\left( q - \tfrac{p_4 + p_1}{2} \right) \tilde{\phi}(p_1) \rra.
\]
In terms of this
\[
	\myint d^d x\ggap e^{-i q \cdot x} \ggap
	\bra{0} \tilde{\phi}(p_4) [\phi(-\sfrac 12 x), \phi(\sfrac 12 x)] \tilde{\phi}(p_1) \ket{0}
    = W(p_4, p_1 | q) - W(p_4, p_1 | \gap{-q}).
\]
Using \Eq{fchoice} for $f(x)$ then gives the momentum space bootstrap equation
\[
\eql{pbootstrap}
\myint dq^0 \bigl[ W(p_4, p_1|q) - W(p_4, p_1| \gap{-q}) \bigr] = 0.
\]

As a check, we verify that \Eq{pbootstrap} is satisfied in generalized free field theory.
Using the explicit form of the 4-point function \eqref{eq:4pt:GFF},
we find
\[
\begin{split}
	W(p_4,p_1| q) - W(p_4,p_1| \gap{-q} )
	&= \frac{(2\pi)^6}{2^{4 \Delta_\phi - 2} \left[ \Gamma(\Delta_\phi) \right]^4}
	(p_1^2)^{\Delta_\phi - 1}
	\delta^2(p_1 + p_4)
	\\
	& \qquad{} \times
	(q^2)^{\Delta_\phi - 1}
	\left[ \Theta(q_+) \Theta(q_-) - \Theta(-q_+) \Theta(-q_-) \right].
\end{split}
\]
This is an odd function of $q^0$, so \Eq{pbootstrap} is satisfied.

For more general choices of $f(x)$, we get a similar equation involving the
convolution of the 4-point function with the Fourier transform of $f(x)$.
This can be used to regularize the integral in \Eq{pbootstrap} when needed.
For instance, when $\Delta_\phi \geq 1$ the integral of each conformal block diverges as $q^0 \to \infty$.
This can be seen with the correlator $\langle TTTT \rangle$ of \Eq{TTTT:W}.
In this case, one can can choose $f(x) = (x^2)^n \delta(x^0)$ for some sufficiently large integer $n$, so that the bootstrap equation becomes
\[
\eql{pbootstrap:regularized}
	\myint dq^0 \, \left( \frac{\partial^2}{\partial q_\mu \partial q^\mu} \right)^n
	\bigl[ W(p_4, p_1|q) - W(p_4, p_1| \gap{-q}) \bigr] = 0.
\]

It would be nice if the bootstrap equation could be restricted to the kinematics
where the momentum space OPE is pointwise convergent.
However, this is not the case for \Eqs{pbootstrap} and \eq{pbootstrap:regularized}, since the integral over
$q$ includes regions where $p$ is arbitrarily large and timelike, which is
outside the region of pointwise convergence.
We expect that this generalizes to other functions $f(x)$.
The reason is that vanishing conditions in position space lead to analyticity
in momentum space, and analytic functions cannot vanish in any finite region.
Nonetheless, the momentum space OPE is expected to converge in the sense
of a distribution.
This means that we can use equations like \Eq{pbootstrap} provided that we
smear the external momenta with smooth test functions.
We leave the investigation of these equations for future work.

One motivation to further study the momentum space bootstrap equation is that
we can kinematically project out the contribution of the identity operator
contribution to the OPE because it contributes only if $p_1 + p_2 = 0$
or $p_1 + p_3 = 0$.
For example, in a reference frame where
$\vec{p}_1 - \pvec{p}_4 = 0$ we can choose
$\vec{q} \ne 0$.
In that case, the integral over $q^0$ in \Eq{pbootstrap} does not include
contributions from the identity operator in either channel.

% =============================================================================
\section{Conclusions}
% =============================================================================

In this paper we have studied the operator product expansion (OPE) of
conformal field theory (CFT) in momentum space, focusing on two spacetime
dimensions.
General principles of quantum field theory imply that there
is an OPE for Wightman functions in Minkowski space that converges
for arbitrary kinematics.
However, this convergence is guaranteed to hold only in the sense of a distribution,
meaning that the  OPE converges only after the correlation functions are smeared
with suitable smooth test functions.
In this paper, we worked out the conformal blocks for this OPE for 2D CFT.
We find that the OPE in fact converges pointwise in a specific
kinematic region, shown in Fig.~\ref{fig:diamond}.
We also formulated a bootstrap equation directly in momentum space that makes
use of this convergent OPE (see for example \Eq{pbootstrap}).

There are a number of important open questions that we leave for future work.
First, it would be interesting to explore whether the momentum space
bootstrap equation can be used to obtain new bounds on CFT data.
This equation involves a convolution of the correlation
functions that requires integration over the region where the correlation
functions do not converge pointwise.
This means that it must be interpreted in the sense
of distributions, and additional smearing with test functions is required
to give equations that can be implemented numerically.
These test functions define the kinematics of the
correlation function.
It would be very interesting to explore the space of kinematics
and see whether there are regions where we can obtain information about
the CFT data, for example using the numerical bootstrap
or extremal functional methods~\cite{Paulos:2019gtx, Mazac:2019shk}.%
\footnote{An interesting feature of the conformal blocks in momentum space
is the presence of double zeroes when the scaling dimension of the exchanged operator
is equal to a double-trace dimension.
Such zeroes are also found in extremal functionals.}
All the tools required for such a study in 2D are provided in this paper.

Another important open question is the generalization of our results to higher
dimensions.
As emphasized in this paper, the conformal blocks in momentum space are
conceptually very simple: they are products of 3-point functions.
This holds in any dimension, even for correlation functions involving
operators with spin.
The obstacles to generalizing this work to higher dimensions are purely technical.
First steps in resolving generic 3-point functions have been taken in
Refs.~\cite{Bautista:2019qxj, Gillioz:2019lgs}, and conformal blocks have been constructed in special cases~\cite{Gillioz:2016jnn, Gillioz:2018kwh, Gillioz:2018mto}.
One interesting question in higher dimensions is whether there is a generalization
of the region of pointwise convergence.
More generally, we would like to have a better understanding of the convergence
of the OPE in momentum space.

We hope that this work will open up new directions in the exploration of
conformal field theory.

\section*{Acknowledgments}
We thank Brian Henning for collaboration in the early stages of this project.
We have benefited from discussions from many other people, including
Denis Karateev,
Marco Meineri,
Anirudha Menon,
Jo\~ao Penedones,
David Simmons-Duffin,
Christopher Verhaaren,
and Matt Walters.
The work of MG at EPFL was supported by the Swiss National Science Foundation through the NCCR SwissMAP.
The work of XL was supported in part by the U.S. Department of Energy under
Grant Number DE-SC0011640.
The work of ML was supported in part by the U.S. Department of Energy
under grant DE-SC-0009999.

\appendices

% =============================================================================
\section{Notation and Conventions}
\label{appsec:Notations}
% =============================================================================

We work in Minkowski space with the `mostly minus' metric $\eta^{\mu\nu} = \diag(+, -, \ldots, -)$. In two dimensions we use lightcone coordinates
\[
x^\pm = x^0 \pm x^1.
\]
in terms of which
\[
x^2 = (x^0)^2 - (x^1)^2 = x^+ x^-.
\]
The metric in lightcone coordinates is
\[
\eta_{+-} = \eta_{-+} = \sfrac 12,
\qquad
\eta_{++} = \eta_{--} = 0.
\]
Fourier transforms are defined by
\[
\tilde{\scr{O}}(p) = \myint d^d x\ggap e^{-i p \cdot x} \ggap \scr{O}(x).
\]
We parameterize 2D momenta using the  lightcone coordinates
$p_\pm = \eta_{\pm\mp} p^\mp = \frac{1}{2} (p^0 \mp p^1)$, so that
\begin{subequations}
\[
p \cdot x &\equiv p_\mu x^\mu = p_+ x^+ + p_- x^-, \\
d^2 x &= \sfrac 12 dx^+ dx^-,
\\
d^2 p &= 2 dp_+ dp_-,
\\
\de^2(p) &= \sfrac 12 \de(p_+) \de(p_-),
\\
\Th(p^0) \Th(p^2) &= \Th(p_+) \Th(p_-),
\]
\end{subequations}
where $\Th$ is the step function.

\subsection{Correlation Functions in Euclidean Position Space}

For 2D conformal field theory we make contact with the Euclidean formulation as follows.
We use
\[
z = x_\text{E}^0 + i x_\text{E}^1,
\qquad
\widebar{z} = x_\text{E}^0 - i x_\text{E}^1.
\]
and label operators by their conformal weights $h$ and $\widebar{h}$, related to the scaling dimension $\Delta$ and spin $s$ by
\[
h = \sfrac 12 (\De + s),
\qquad\qquad
\widebar{h} = \sfrac 12 (\De - s).
\]
2- and 3-point correlation functions are fixed by conformal symmetry and take the form
\[
% \bra{0} \O(x_{2\text{E}}) \O(x_{1\text{E}}) \ket{0}
\avg{ \O(z_2) \O(z_1) }
% \bra{0} \O(z_2) \O(z_1) \ket{0}
&= \frac{1}{(z_{21})^{2h} (\widebar{z}_{21})^{2\widebar{h}}},
\eql{2ptapp}
\\
\avg{ \O_3(z_3) \O_2(z_2) \O_1(z_1) }
% \bra{0} \O_3(x_{3E}) \O_2(x_{2E}) \O_1(x_{1E}) \ket{0}
&= \frac{\la_{321}}
{(z_{21})^{h_{12|3}} (z_{31})^{h_{13|2}} (z_{32})^{h_{23|1}}
(\widebar{z}_{21})^{\widebar{h}_{12|3}} (\widebar{z}_{31})^{\widebar{h}_{13|2}}
(\widebar{z}_{32})^{\widebar{h}_{23|1}}},
\eql{3ptapp}
\]
where $\lambda_{321}$ is an OPE coefficient,
$z_{ab} = z_a - z_b$ and $h_{ab|c} = h_a + h_b - h_c$.
The 4-point correlation functions can be parametrized as
\[
\begin{split}
	\bra{0} \O_4(z_4) \O_3(z_3) \O_2(z_2) &\O_1(z_1) \ket{0}
	= \frac{G(\eta, \widebar{\eta})}
	{(z_{43})^{h_4 + h_3} (z_{21})^{h_2 + h_1}
	(\widebar{z}_{43})^{\widebar{h}_4 + \widebar{h}_3} (z_{21})^{\widebar{h}_2 + \widebar{h}_1}}
	\\
	&\qquad\quad{} \times
	\left( \frac{z_{41}}{z_{42}} \right)^{h_2 - h_1}
	\left( \frac{z_{41}}{z_{31}} \right)^{h_3 - h_4}
	\left( \frac{\widebar{z}_{41}}{\widebar{z}_{42}} \right)^{\widebar{h}_2 - \widebar{h}_1}
	\left( \frac{\widebar{z}_{41}}{\widebar{z}_{31}} \right)^{\widebar{h}_3 - \widebar{h}_4}
\eql{4ptxapp}
\end{split}
\]
in terms of a function $G$ of the cross-ratios
\[
	\eta = \frac{z_{43} z_{21}}{z_{42} z_{31}},
	\qquad\qquad
	\widebar{\eta} = \frac{\widebar{z}_{43} \widebar{z}_{21}}{\widebar{z}_{42} \widebar{z}_{31}}.
\]
This function $G$ admits the expansion%
\[
	G(\eta, \widebar{\eta}) = \sum_{\psi} \lambda_{43\psi} \lambda_{\psi21}
	G_{h_\psi}(\eta) G_{\widebar{h}_\psi}(\widebar{\eta})
	\eql{PSBlockExpansion}
\]
where $G_h$ are the conformal blocks~\cite{Dolan:2000ut}
\[
	G_h(\eta) = \eta^h {}_2F_1\left( h - h_4 + h_3, h + h_2 - h_1; 2h; \eta \right).
	\label{eqn:PSBlock}
\]
Note that for each operator $\psi$ with scaling dimension $\Delta$ and spin $s \neq 0$, there is another operator with identical scaling dimension and opposite spin $-s$ by CPT symmetry: one has conformal weights
$(h, \widebar{h}) = \left( \frac{\Delta + s}{2}, \frac{\Delta - s}{2} \right)$
and the other $\left( \frac{\Delta - s}{2}, \frac{\Delta + s}{2} \right)$.
In this way the sum in \Eq{PSBlockExpansion} can be viewed as a double sum over conformal weights, without restrictions regarding the relative size of  $h$ and $\widebar{h}$.

\subsection{Analytic Continuation from Euclidean to Minkowski Space}
\label{appsec:Continuation}

Coordinates in 2-dimensional Minkowski space are related to Euclidean ones by
\[
\eql{analyticcont}
x_\text{E}^0 = i x^0,
\qquad
x_\text{E}^1 = x^1.
\]
Under this analytic continuation, the 2- and 3-point functions \eqref{eq:2ptapp} and \eqref{eq:3ptapp} become
\[
\bra{0} \O(x_2) \O(x_1) \ket{0}
&= \frac{(e^{-i \pi})^{h + \widebar{h}}}{(x_{21}^+ - i\ep)^{2h} (x_{21}^- - i\ep)^{2\widebar{h}}},
\eql{2ptxapp}
\\
\begin{split}
\bra{0} \O_3(x_3) \O_2(x_2) \O_1(x_1) \ket{0}
&= \la_{321} \frac{(e^{-i\pi/2})^{h_1 + h_2 + h_3}}
{(x_{21}^+-i\ep)^{h_{12|3}} (x_{31}^+-i\ep)^{h_{13|2}} (x_{32}^+-i\ep)^{h_{23|1}}}
\\
&\qquad\ {} \times
\frac{(e^{-i\pi/2})^{\widebar{h}_1 + \widebar{h}_2 + \widebar{h}_3}}
{(x_{21}^- -i\ep)^{\widebar{h}_{12|3}} (x_{31}^- -i\ep)^{\widebar{h}_{13|2}}
(x_{32}^- -i\ep)^{\widebar{h}_{23|1}}}.
\eql{3ptxapp}
\end{split}
\]
This gives the standard $i\ep$ prescription for Wightman functions.
It can be understood from the fact that the Wightman functions
$\bra{0} \O_n(x_n) \cdots \O_1(x_1) \ket{0}$
are analytic in the region where
$\Im x_{ij}^0 < 0$ for $i > j$ by positivity of energy.

It is worth checking that the phases on the right-hand sides of \Eqs{2ptxapp} and \eq{3ptxapp}
are compatible with reflection positivity in Euclidean space.
For a 2-point function, we have
\[
\bra{0} \O(x^0 = -i\tau, x^1 = 0) \O(0) \ket{0} = \bra{0} \O(0) e^{-H \tau} \O(0) \ket{0}
= \braket\Psi\Psi > 0,
\eql{reflectpos2pt}
\]
where $H$ is the Hamiltonian and $\ket\Psi = e^{-H \tau/2} \O\ket{0}$.
Note we require $\tau > 0$ (time ordering) for this to make sense.
We now evaluate \Eq{2ptxapp} by analytic continuation:
\[
x^0 = \lim_{\th \, \to \, \frac\pi{2}} e^{-i\th} \tau,
\qquad
x^1 = 0.
\]
This gives
\[
\bra{0} \O(-i\tau) \O(0) \ket{0} = \frac{(e^{-i\pi})^{h + \widebar{h}}}
{(\tau e^{-i\pi/2})^{2h}
(\tau e^{-i\pi/2})^{2\widebar{h}}} > 0,
\]
in agreement with \Eq{reflectpos2pt}.

% =============================================================================
\section{Asymptotic Behavior of Hypergeometric Functions}
\label{appsec:Asymptotics}
% =============================================================================

In order to compute the limit~\eqref{eq:Wasymptotics} of the holomorphic conformal block, we use the asymptotic expansions of the hypergeometric functions appearing in \Eq{V3} under $h \to \infty$.
Working specifically with $0<z<1$, we obtain
\begin{equation}
	{}_2F_1\left(h+a, h+b; 2h; z \right)
	\stackrel{h \to \infty}{\simeq}
	(1-z)^{-(a+b)/2-1/4} \left( \frac{2}{1 + \sqrt{1-z}} \right)^{2h-1}
	\label{eqn:2F1Asymptotic1}
\end{equation}
and
\begin{align}
\begin{split}
\!\!\!\!
	& {}_2F_1\left( 1 + a - h, a + h; 2b; z \right)
	\\
	& \quad
	\stackrel{h \to \infty}{\simeq}
	\frac{\Gamma(2b)}{\sqrt{\pi}} h^{1/2 - 2b} z^{1/4 - b} (1-z)^{b - a - 3/4}
	\Re \left[ e^{i \pi (b - 1/4)} \left( 1 - 2z - 2i \sqrt{z(1-z)} \right)^{h-1/2} \right]
	\label{eqn:2F1Asymptotic2}
\end{split}
\end{align}
Note that the term in square bracket has modulus one, so its real part is bounded in the interval $[-1,1]$.

% =============================================================================
\section{OPE Convergence in Generalized Free Scalar Theory}
\label{appsec:GFF}
% =============================================================================

We detail in this appendix the results of Section~\ref{sec:convergence:GFF} for the 4-point function of a scalar operator $\phi$ in generalized free field theory. We show in particular that the OPE converges to the expected result, but only in a distributional sense.

The operators that contribute to the expansion have conformal weights
\[
	h = \Delta_\phi + m,
	\qquad\qquad
	\widebar{h} = \Delta_\phi + \widebar{m},
	\qquad\qquad
	m, \widebar{m} \in \mathbb{N}.
\]
Plugging these values in the definition \eqref{eq:HolomorphicConformalBlocks} of the holomorphic conformal blocks, one gets
\[
	W_{\Delta_\phi+m}(k_4, k_3, k_2, k_1) &=
	\frac{(2\pi)^3}{2 \sqrt{2}}
	(k_1 k_2 k_3 k_4)^{(\Delta_\phi - 1)/2} (k_1 + k_2)^{-1}
	\Theta(k_2) \Theta(-k_3)
	\nn
	& \quad \times
	\frac{\Gamma(2\Delta_\phi + 2m)}{\Gamma(\Delta_\phi + m)^2}
	P_{\Delta_\phi + m - 1}^{1 - \Delta_\phi}\left( \frac{k_2 - k_1}{k_1 + k_2} \right)
	P_{\Delta_\phi + m - 1}^{1 - \Delta_\phi}\left( \frac{k_3 - k_4}{k_3 + k_4} \right)
\]
where $P_\lambda^\mu$ is the associated Legendre function.%
\footnote{The associated Legendre function is a special case of the hypergeometric function:
$$
	P_{\Delta_\phi + m - 1}^{1 - \Delta_\phi}\left( \frac{k_2 - k_1}{k_1 + k_2} \right)
	= \frac{1}{\Gamma(\Delta_\phi)} \left( \frac{k_2}{k_1} \right)^{(1 - \Delta_\phi)/2}
	{}_2{F_1}\left( 1 - \Delta_\phi - m, \Delta_\phi + m; \Delta_\phi; \frac{k_1}{k_1 + k_2} \right).
$$}
As indicated by the $\Theta$-functions, the conformal block vanishes when $k_2 < 0$ or $k_3 > 0$.

These blocks can be combined with the known OPE coefficients~\cite{Fitzpatrick:2011dm}
\[
\eql{GFFOPE}
	\lambda _{m, \widebar{m}}^2
	= \left[ {1 + {{\left( { - 1} \right)}^{m + \bar m}}} \right]
	\frac{{\Gamma \left( {2{\Delta _\phi } + m - 1} \right)\Gamma \left( {2{\Delta _\phi } + \bar m - 1} \right){{\left[ {\Gamma \left( {{\Delta _\phi } + m} \right)\Gamma \left( {{\Delta _\phi } + \bar m} \right)} \right]}^2}}}
	{{m! \widebar{m}! \Gamma \left( {2{\Delta _\phi } + 2m - 1} \right)\Gamma \left( {2{\Delta _\phi } + 2\bar m - 1} \right){{\left[ {\Gamma \left( {{\Delta _\phi }} \right)} \right]}^4}}}
\]
to write the expansion as
\[
\begin{split}
	\lla \tilde{\phi}(p_4) \tilde{\phi}(p_3) \tilde{\phi}(p_2) \tilde{\phi}(p_1) \rra
	&= \frac{(2\pi)^6}{2^{4\Delta_\phi - 2} \left[ \Gamma(\Delta_\phi) \right]^4}
	(p_4^2 p_3^2 p_2^2 p_1^2)^{(\Delta_\phi - 1)/2}
	\\
	& \quad \times
	\left[ \delta^2(p_1 + p_2) +
	2 \, \frac{f_+(p_{i+}) f_+(p_{i-}) + f_-(p_{i+}) f_-(p_{i-})}{(p_1 + p_2)^2} \right],
\eql{GFFexpansion:f}
\end{split}
\]
where we have denoted
\[
\begin{split}
	f_\pm(k_i) = \sum_{m = 0}^\infty (\pm 1)^m &
	\frac{(2\Delta_\phi + 2m - 1) \Gamma(2\Delta_\phi + m - 1)}{m!}
	\\
	&{}\times
	P_{\Delta_\phi + m - 1}^{1 - \Delta_\phi}\left( \tfrac{k_2 - k_1}{k_1 + k_2} \right)
	P_{\Delta_\phi + m - 1}^{1 - \Delta_\phi}\left( \tfrac{k_3 - k_4}{k_3 + k_4} \right).
\end{split}
\]
As emphasized in Section~\ref{sec:convergence:GFF}, this series does not converge in a pointwise manner. However, it converges in a distributional sense to the delta function, by use of the identity
\[
	\sum_{m = 0}^\infty (\pm 1)^m
	\frac{(2\Delta_\phi + 2m - 1) \Gamma(2\Delta_\phi + m - 1)}{m!}
	P_{\Delta_\phi + m - 1}^{1 - \Delta_\phi}(x)
	P_{\Delta_\phi + m - 1}^{1 - \Delta_\phi}(y)
	= 2 \delta(x \pm y),
\]
so that we have
\[
	f_+(k_i) = (k_1 + k_2) \delta(k_1 + k_3),
	\qquad\qquad
	f_-(k_i) = (k_1 + k_2) \delta(k_1 + k_4).
\]
Using this in Eq.~\eqref{eq:GFFexpansion:f} precisely reproduces Eq.~\eqref{eq:4pt:GFF}.

\newpage
\frenchspacing
\bibliographystyle{utphys}
\bibliography{mycites}

\end{document}